\documentstyle[12pt]{article}

\begin{document}

\thispagestyle{empty}

\hfill \parbox{45mm}{{ECT*-98-012} \par June 1998} 

\vspace*{15mm}

\begin{center}
{\LARGE Tunneling of a Massless Field}

\medskip

{\LARGE through a 3D Gaussian Barrier.}

\vspace{22mm}

{\large Giovanni Modanese}
\footnote{E-mail: {\tt modanese@science.unitn.it}.}

\medskip

{\em European Centre for Theoretical Studies in Nuclear Physics and 
Related Areas \par Villa Tambosi, Strada delle Tabarelle 286 \par
I-38050 Villazzano (TN) - Italy}

\bigskip \bigskip

\end{center}

\vspace*{10mm}

\begin{abstract}

We propose a method for the approximate computation of the Green function
of a scalar massless field $\phi$ subjected to potential barriers of given
size and shape in spacetime. This technique is applied to the case of a 3D
gaussian ellipsoid-like barrier, placed on the axis between two pointlike
sources of the field. Instead of the Green function we compute its
temporal integral, that gives the static potential energy of the
interaction of the two sources. Such interaction takes place in part by
tunneling of the quanta of $\phi$ across the barrier. We evaluate
numerically the correction to the potential in dependence on the barrier
size and on the barrier-sources distance. 


\end{abstract}

\newpage

\noindent
{\Large \bf I. Introduction.}

\bigskip

In Quantum Field Theory it is useful in several occasions to have a
general expression for the Euclidean two-point correlation function of a
massless scalar field $\phi$ in the presence of potential ``barriers" in
spacetime of the form
	\begin{equation}
	V(\phi(x)) = \xi J_\Omega(x) \left[ 
	\phi^2(x) - \phi^2_0 \right]^2,
\label{e11}
\end{equation}
	$J_\Omega(x)$ being the characteristic function of the 4-region
$\Omega$ where the potential has support ($J_\Omega=1$ for $x \, \epsilon
\, \Omega$, $J_\Omega=0$ elsewhere). The region $\Omega$ can be multiple
connected, thus representing several barriers placed at different points
in spacetime. 

Possible applications are connected for instance to the fact that a
potential of the form (\ref{e11}) represents a localized imaginary mass
term ($m^2<0$) in the action of the scalar field $\phi$.  Terms of this
kind can be present in cosmological models with inflationary fields. It is
also known that every quantum field with non-vanishing vacuum expectation
value (VEV) has a global imaginary mass term in its lagrangian \cite{wei},
which couples to the gravitational field as a cosmological term; one can
show \cite{cc} that if the VEV is not constant but depends on $x$, it
becomes a {\it local} cosmological term for the gravitational field. 

More generally, suppose we have a system of two interacting fields and
regard one of them (or its VEV) as a fixed external source. The coupling
term of the two fields becomes a local constraint for the dynamical field,
a sort of external potential localized in the regions where the external
field has support. It is therefore important to study the tunneling of the
dynamical field through these regions, that is, its Green functions.  Note
that in systems like this translational invariance is generally lost. 

It is easy to check that the potential $V(\phi)$ in eq.\ (\ref{e11}) 
implements in fact a constraint in the functional integral of the field: 
writing this integral as
	\begin{displaymath}
	z = \int d[\phi] \, \exp \left[-\int d^4x \, (\partial \phi)^2 
	- \int d^4x \, V(\phi) \right],
\end{displaymath}
	one sees that for large $\xi$ the square of the field is forced to
take the value $\phi_0^2$ within the region $\Omega$. 

For a characteristic function $J_\Omega(x)$ like the one specified above
we say that the constraint is imposed in a ``sharp"  way: in spacetime the
potential barrier looks like a step at the boundary of $\Omega$. Smoothing
$J_\Omega$ we can obtain a smooth potential barrier. In the following we
shall be more interested in this second case. 

Note that the potential (\ref{e11}) has the shape of a double {\it well},
as long as considered only a function of the field $\phi$, but regarded as
a function of $x$ it is positive and reminds much more a {\it barrier}.

Let us focus on the case of weak fields, such that $\phi^4$ can be
disregarded with respect to $\phi^2$. If the product $\gamma \equiv \xi
\phi_0^2$ is small, then the effect of the barriers on field correlations
is small, too, and can be treated as a perturbation. One can solve the
equation for the modified propagator $G'(x_1,x_2)=\langle
\phi(x_1)\phi(x_2)\rangle_V$ in closed form (see the Appendix), finding
that $G'$ is given by a double inverse Fourier transform, with the direct
transform of $J_\Omega$ evaluated at $(p+k)$:
	\begin{eqnarray}
	& & G(x_1,x_2) = G^0(x_1,x_2) + \gamma G'(x_1,x_2); \nonumber \\
	& & G^0(x_1,x_2) = \int d^4k \, e^{-ik(x_1-x_2)}; \nonumber \\
	& & G'(x_1,x_2) = \int d^4p \int d^4k \, e^{ipx_1} e^{ikx_2} 
	\frac{\tilde{J}_\Omega(p+k)}{k^2 p^2}.
\label{e13}
\end{eqnarray}

	In finite-dimensional quantum mechanics computing $G'(x_1,x_2)$
corresponds to compute the Feynman transition amplitude, related in turn
to the system's wavefunction in the presence of barriers. In field theory
the intuitive meaning of $G'(x_1,x_2)$ is less immediate.  However, we can
derive from $G'(x_1,x_2)$ a quantity with a direct physical
interpretation: the static potential $U({\bf x}_1,{\bf x}_2)$ of the
interaction of two pointlike sources $q_1$ and $q_2$ of the field $\phi$
at rest. This interaction is mediated by the exchange of quanta of $\phi$.
If the barriers are placed somewhere between the sources, the interaction
is clearly affected, but it still takes place -- provided the product
$\gamma$ is small -- with the quanta of $\phi$ ``tunneling"  through the
barriers (or passing over the wells, depending on the interpretation).

The leading contribution to the static potential $U({\bf x}_1,{\bf x}_2)$
is obtained from (\ref{e13}) as follows \cite{sta}. First one defines
$J_\Omega(x)$ as the product of a 3D function $j_\Omega({\bf x})$ and a
function constant in time, then one integrates over $t_1$ and $t_2$,
multiplies by $q_1q_2$ and divides by $-T$, taking the limit for $T \to
\infty$.  The result is
	\begin{eqnarray}
	U({\bf x}_1,{\bf x}_2) & = & U^0({\bf x}_1,{\bf x}_2) 
	+ \gamma U'({\bf x}_1,{\bf x}_2) = \nonumber \\
	& = & \frac{q_1 q_2}{|{\bf x}_1-{\bf x}_2|} - 
	\gamma (2\pi)^8 q_1 q_2 \int d{\bf p} \int d{\bf k} \, 
	e^{i{\bf px}_1} e^{i{\bf kx}_2} \, 
	\frac{\tilde{j}_\Omega({\bf p}+{\bf k})}{{\bf k}^2 {\bf p}^2}.
\label{e14}
\end{eqnarray}
	This formula is easily generalized to the case of $N$ charges
$q_1,...,q_N$, placed respectively at ${\bf x}_1,...,{\bf x}_N$. 

A limit case of the physical situation we are considering is represented
by the electrostatic potential of pointlike charges in the presence of
perfect conductors. In this case the field is exactly zero within the
region $\Omega$, and $\Omega$ has sharp boundaries --- thus $j_\Omega({\bf
x})$ is a step function and $\tilde{j}_\Omega({\bf p})$ a strongly
oscillating function. Eq.\ (\ref{e14}) could be applied to this case only
if the parameters $\phi_0$ and $\xi$ could be chosen in such a way that
$\phi_0 \to 0$ and $\xi \to \infty$, the product $\gamma=\xi \phi_0^2$
still being finite and small.  We know, however, that usually in an
electrostatic system the change in potential energy due to the presence of
perfect conductors is not just a small correction. (It can be computed
exactly, in principle, solving a classical field equation with suitable
boundary conditions.)

The case of interest here is actually more subtle. In the following
$j_\Omega({\bf x})$ is supposed to be a smooth function and both $\phi_0$
and $\xi$ are taken to be finite. The field square has only a certain
probability to be equal to $\phi_0^2$ within $\Omega$. This probability is
maximum at the center of $\Omega$ and decreases towards the boundary of
$\Omega$.  Since $j_\Omega({\bf x})$ is smooth (a gaussian function), its
Fourier transform $\tilde{j}_\Omega({\bf p})$ is smooth too, and the
integral (\ref{e14}) can be computed numerically.

It is interesting to study $U'$ in dependence on the geometrical features
of the barrier $\Omega$ and on the position of $q_1$ and $q_2$ with
respect to it. Take, for instance, a finite size barrier (gaussian
ellipsoid, see Section II) lying on the axis joining ${\bf x}_1$ to ${\bf
x}_2$. We may expect that if one of the two charges is close to $\Omega$,
then $|U'/U^0|$ is larger, decreasing if both charges are far away from
$\Omega$ --- or if $\Omega$ is not on their axis. This behavior is
confirmed and specified by our numerical results. 

The paper is organized as follows. In Section II we compute the leading
correction to the static potential for a barrier with the shape of an
ellipsoid. Due to the peculiar behavior of the integrand, the procedure
for numerical integration is not trivial and requires some care. We
describe it in detail. Results are given in Section III. They concern in
particular the dependence of the correction to $U({\bf x}_1,{\bf x}_2)$ on
the geometrical setting (size of the barrier and its position with respect
to the pointlike sources). Far from exploring all the conceivable
variations and related phenomenology, the main aim of this work is to show
that the general technique can be successfully applied to real cases.

\vskip 2 cm

\noindent
{\Large \bf II. The case of two static sources.}

\bigskip

Let us focus now on a configuration with two static sources and one
barrier only. We choose our reference frame in such a way that the sources
lie on the $z$-axis: 
	\begin{displaymath}
	   {\bf x}_1 = (0,0,L_1); \ \ \ \  {\bf x}_2 = (0,0,-L_2).
\end{displaymath}
The spatial shape and size of the barrier are defined by the function
\begin{equation}
	j_\Omega({\bf x}) = \exp \left(-\frac{x^2+y^2}{a^2} 
	- \frac{z^2}{b^2} \right).
\label{e21a}
\end{equation}
	This means that the region $\Omega$ is like an ellipsoid centered
at the origin, with simmetry axis along $Oz$, radius of the order of $a$
and thickness of the order of $b$. We suppose that $a>b$, thus the
ellipsoid is ``squeezed" on the $xy$-plane.  More precisely, the region
$\Omega$ itself is not sharply defined, but the surfaces where
$j_\Omega({\bf x})$ is constant are ellipsoids. For instance, on the
surface defined by
	\begin{displaymath}
	\frac{x^2+y^2}{a^2} + \frac{z^2}{b^2} = 1
\label{e21b}
\end{displaymath}
	the function $j_\Omega({\bf x})$ is constant and equal to $e^{-1}$. 

The Fourier transform of (\ref{e21a}) is
	\begin{equation}
	\tilde{j}_\Omega({\bf p}) = \pi^{3/2} a^2 b \exp
	\left\{ \frac{1}{4} \left[ -a^2 (p_x^2 + p_y^2) - b^2 p_z^2
	\right] \right\}.
\label{e21c}
\end{equation}

The charges $q_1$ and $q_2$ can be taken to be unitary and the distances
$L_1$ and $L_2$ expressed as multiples of the ellipsoid radius $a$:  $L_1
\equiv n_1a, \ L_2 \equiv n_2a$. From (\ref{e14}), (\ref{e21c}) we obtain

	\begin{displaymath}
	U'({\bf x}_1,{\bf x}_2) = - (2\pi)^8 \pi^{3/2} a^2 b \int d{\bf p} 
	\int d{\bf k} \, \frac{e^{ik_z n_1 a - ip_z n_2 a}}
	{{\bf k}^2 {\bf p}^2}
	\exp \left[-\frac{1}{4} a^2({\bf p}+{\bf k})_{xy}^2- 
	\frac{1}{4} b^2(p_z+k_z)^2 \right],
\end{displaymath}

\noindent 
	where ${\bf V}_{xy}$ denotes the component of a vector ${\bf V}$
in the plane $xy$.  In the following we shall be most interested in the
case with one charge far from the barrier ($n_1\gg 1$), while the other
charge is close to it (typically in our numerical calculations $n_2$
ranges between 1 and 15). Accordingly we set $n_1^{-1}=\varepsilon$,
$n_2=n$.  After rescaling $k_z \to \varepsilon k_z$ we obtain

	\begin{displaymath}
	U'({\bf x}_1,{\bf x}_2) = -(2\pi)^8 \pi^{3/2} \frac{a^3 b}{L_1} 
	\int d{\bf p} \int d{\bf k} \, \frac{e^{ik_z a - ip_z n a}}
	{{\bf p}^2(k_x^2+k_y^2+\varepsilon k_z^2)}
	\exp \left[ -\frac{1}{4} a^2({\bf p}+{\bf k})_{xy}^2-
	\frac{1}{4} b^2(p_z+\varepsilon k_z)^2 \right].
\end{displaymath}

	Then we eliminate any further dimensional parameters by
rescaling $k \to 2k/a$ and $p \to 2p/a$, obtaining

	\begin{displaymath}
	U'({\bf x}_1,{\bf x}_2) = -(2\pi)^8 \pi^{3/2} \frac{4ab}{L_1} 
	\int d{\bf p} \int d{\bf k} \, \frac{e^{ik_z - ip_z n}}
	{{\bf p}^2(k_x^2+k_y^2+\varepsilon k_z^2)}
	\exp \left[ -({\bf p}+{\bf k})_{xy}^2-
	\rho^2(p_z+\varepsilon k_z)^2 \right],
\end{displaymath}

\noindent 
	where $\rho=b/a$ is the ratio between the thickness $b$ and the
radius $a$ of the ellipsoid. 

Next we introduce the polar variables $\theta_k$, $\theta_p$, $\phi_k$,
and $\phi_p$. In the following $k$ and $p$ will not denote four-vectors
anymore, but $|{\bf k}|$ and $|{\bf p}|$, respectively. The square of the 
component of the vector $({\bf p}+{\bf k})$ in the $xy$ plane is
	\begin{displaymath}
	({\bf p}+{\bf k})_{xy}^2 = p^2 \sin^2 \theta_p + 
	k^2 \sin^2 \theta_k + 2pk \sin \theta_p \sin \theta_k 
	\cos(\phi_k-\phi_p).
\end{displaymath}
The other components are
\begin{eqnarray}
	& & k_z = k \cos \theta_k; \ \ \ \  p_z = p \cos \theta_p; \nonumber 
\\ 
	& & k_x^2+k_y^2 = {\bf k}_{xy}^2 = k^2 \sin^2 \theta_k.
\nonumber
\end{eqnarray}
Finally, introducing the variables
\begin{displaymath}
	s = \cos \theta_k, \ \ \  t = \cos \theta_p, \ \ \     
	\phi = (\phi_k-\phi_p)
\end{displaymath}
one obtains, remembering that the integrand is even in $s$, $t$, the 
following basic formula:
\begin{eqnarray}
	U'({\bf x}_1,{\bf x}_2) &=& -\frac{(2\pi)^{10}}{\sqrt{\pi}}
	\, \frac{ab}{L_1} \, 2\pi 
	\int_0^{2\pi} d \phi \int_{-1}^1 ds \int_{-1}^1 dt 
	\int_0^\infty dk \int_0^\infty dp \, \frac{\cos(ks-npt)}
	{1-s^2(1-\varepsilon^2)} \times \nonumber \\
	& & \ \times 
	\exp \left[ -\rho^2(pt+\varepsilon ks)^2 - p^2(1-t^2) 
	- k^2(1-s^2) - 2 pk \cos \phi \sqrt{(1-t^2)(1-s^2)} \right]
	\nonumber \\
	& \equiv & -\frac{(2\pi)^{11}}{\sqrt{\pi}} \, \frac{ab}{L_1}  
	\int_0^{2\pi} d \phi \int_{-1}^1 ds \int_{-1}^1 dt 
	\int_0^\infty dk \int_0^\infty dp \,
	f(\phi,s,t,k,p;\varepsilon,\rho,n)  \nonumber \\
	& \equiv & -\frac{(2\pi)^{11}}{\sqrt{\pi}} \, \frac{ab}{L_1} 
	F(\varepsilon,\rho,n).
\label{e28}
\end{eqnarray}

\vskip 1 cm
\noindent
{\large \bf A. Preliminary study of the integrand.}

\medskip

It is important to discuss in advance the case in which $\rho$ and
$\varepsilon$ take values much smaller than 1, that is, $\Omega$ is very
thin and the distance of the first charge from $\Omega$ is much larger
than $a$. When $t$ and $s$ approach +1 or -1, for small values of $\rho$
the integral over $k$ and $p$ converges very slowly at infinity and the
factor $\cos(ks-npt)$ performs a large number of oscillations. For very
small $\varepsilon$ there are many more oscillations in $k$ than in $p$.
(In the limit $\rho \to 0$ the integral makes sense only as a
distribution. We shall never approach this limit, however.) 

Let us set, for instance, $s=1$, $t=1$ and $\phi=\pi/2$ in the argument of 
the exponential in (\ref{e28}). We obtain the exponential factors
	\begin{equation}
	\exp \left[ -\rho^2(pt+\varepsilon ks)^2 \right] = 
	\exp \left[ -(\rho p)^2\right] \, \exp 
	\left[ -(\rho \varepsilon k)^2 \right] 
	\, \exp \left[-2 \rho^2 \varepsilon kp\right]. 
\label{uffa}
\end{equation} 
	The first factor on the r.h.s.\ of (\ref{uffa}) has a range in $p$
of the order of $\rho^{-1}$ and the second factor has a range in $k$ of
the order of $(\rho \varepsilon)^{-1}$. The third factor has a range in
$p$, for fixed $k$, of the order of $(\rho \sqrt{\varepsilon} k)^{-1}$ and
a range in $k$, for fixed $p$, of the order of $(\rho \sqrt{\varepsilon}
p)^{-1}$.  Fortunately this latter factor is not relevant: if its range is
larger than the other two ranges then it does not play any role;  if it is
smaller then it is sufficient to refer to the other ranges.

As soon as $s^2$ and $t^2$ go away from 1, the number of oscillations of
the integrand decreases. For instance, setting $s=t=0.98$ we obtain the 
exponential factors
	\begin{displaymath}
	\sim \exp \left[ -(\rho p)^2 - (\rho \varepsilon k)^2 
	- 2 \rho^2 \varepsilon kp \right] \, \exp \left[- 
	0.04 \, p^2 - 0.04 \, k^2 \right].
\end{displaymath}
	When $\rho$ is much smaller than 1 the range of this product is
determined by the second exponential and does not depend on $\rho$. 

It is also easy to take into account the term proportional to $\cos\phi$.
After setting $\phi=\pi$ that term gives a positive contribution to the
argument of the exponential; thus studying the range of the resulting
expression we obtain an upper limit valid for any $\phi$. 

\vskip 1 cm
\noindent
{\large \bf B. Integration domains.}

\medskip

Independently of the considerations above, it is possible to plot the
integrand $f(\phi,s,t,k,p;\varepsilon,\rho,n)$ for several different
values of $\rho$ and $\varepsilon$ and check the ranges of the
exponentials. In order to better control the oscillations of $f$,
we study
it in 4 different domains of the variables $s$, $t$: 

\medskip

Domain 1: $t,s \, \epsilon \, [0,1-\alpha]$;

Domain 2: $t \, \epsilon \, [0, 1-\alpha]; \ s \, \epsilon \, [1-\alpha, 1]$;

Domain 3: $t \, \epsilon \, [1-\alpha, 1]; \ s \, \epsilon \, [0, 1-\alpha]$;

Domain 4: $t,s \, \epsilon \, [1-\alpha, 1-\alpha]$.

\medskip

\noindent 
	A typical value of $\alpha$ employed in the program is
$\alpha=0.02$. The total integration domain in $s$, $t$ is obtained by
``reflecting" each of the domains above with respect to one axis and then
reflecting again the result with respect to the origin ($s \to - s$, $t
\to -t$, $s,t \to -s,-t$).  In each domain $i$ there is a maximum value
for the variables $k$ and $p$, beyond which $f$ is equal to zero for any
practical purpose.  Denoting by $K_i$ and $P_i$ these ranges, Table 1
shows the results found for some considered values of $\varepsilon$ and
$\rho$.

Since the integration over $k$ and $p$ is extended to wide ranges,
the most reasonable technique for the numerical computation of the
integral (\ref{e28}) appears to be a Montecarlo sampling of the integrand.
The sampling algorithm evaluates the average value of $f$ in each domain,
extending the values of $k$ and $p$ up to the maximum range necessary for
that domain. At the end the global average is computed, weighing each
single average with the ratio between the domain volume and the
total volume. Denoting by $f_i$ the average of $f$ in the domain $i$ and
by $V_i$ the domain volume we have
	\begin{equation}
	F = \sum_{i=1}^4  f_i V_i = 8\pi \left[(1-\alpha)^2 K_1 P_1 f_1+ 
	\alpha(1-\alpha) K_2 P_2 f_2 + \alpha(1-\alpha) K_3 P_3 f_3 + 
	\alpha^2 K_4 P_4 f_4 \right].
\label{e211}
\end{equation}

\vskip 2 cm

\noindent
{\Large \bf III. Results of the numerical integration.}

\bigskip

The contributions of the Domains 2, 3 and 4 to the integral $F$ (compare
(\ref{e211})) are found to be small with respect to the contribution of
Domain 1. The fluctuations of the average of $f$ in Domains 2 and 4 (where
$s^2$ approaches 1) may be very large. In order to achieve a sufficient
precision these regions have been sampled with a large number of points
(up to $\sim 10^{10}$). The standard routine ``{\tt ran2}" \cite{f77} was 
used for random numbers generation. 

The dependence of the integral $F$ on the parameters $\varepsilon$ and
$\rho$ is very weak, thus $U'$ depends on $a$, $b$ and $L_1$ mainly as
$ab/L_1$ (see eq.\ (\ref{e28})). The study of the dependence of $U'$ on
$n$ is more difficult, because this dependence is entirely contained in
the integral $F$ and can be evaluated only numerically. One needs to
insert in the program a cycle which samples the integrand for different
values of $n$, typically between 1 and 15. This is possible because the
ranges $P_i$, $K_i$ do not depend on $n$. 

The numerical evaluation of $F$ as a function of $n$ in the range
$n=1...15$, with $\varepsilon=0.1$ and $\rho=0.3$, gives the results shown
in Fig.\ 1. With $\rho=0.1$ and $\rho=0.032$ one obtains very similar
results, thus confirming the weak dependence on $\rho$ (Fig.\ 2). As
expected varying $\varepsilon$ does not affect much the value of $F$
either, since the dependence on the distance $L_1$ is already factorized
out of the integral (compare Fig.\ 3).

Fig.s 1, 2, 3 reveal an exponential behavior of $F(n)$ of the form
	\begin{displaymath}
	F(n) \sim \exp(-mn+q) + b.
\end{displaymath}
	It is also clear just from the graphs that the exponential
decrease of $F$ for large $n$ leaves an asymptotic value $F=b$, with $b$
in the interval 0.1-0.2. This is an interesting behavior, as it means that
the ``shadow" produced by the barrier in the static field of the two
sources has a long, constant tail. 

A least-squares fit of the data gives the results of Table 2. Excluding
from the fit the first two points ($n=1,2$) we obtain better estimates for
the distribution tail and for $b$. The errors on the parameters of the
fit, in particular those on $b$, are small. They can be estimated knowing
that the least-squares sum of the percentual errors $S=\sum_n\{1-[
\exp(-mn+q)+b]/F(n)\}^2$ has a minimum value $S_{min}\sim 0.05$ and that
its second partial derivatives at the minimum are of the order of
$\frac{\partial^2 S}{\partial b^2} \sim 4 \cdot 10^2$, $\frac{\partial^2
S}{\partial m^2} \sim 10^2$, $\frac{\partial^2 S}{\partial q^2} \sim 10$.

\vskip 2 cm

\noindent
{\Large \bf IV. Conclusions.}

\bigskip

Our technique for the computation of the Green function and the static
potential of two pointlike sources appears to work well for weak fields,
yielding reasonable results. The method is based upon a double 3D Fourier
transform of the function which represents size and position in space
of the potential well or barrier. This double transform is necessary, due
to the lack of translational invariance of the system. Its numerical
evaluation requires a preliminary analytical study and a subdivision of
the integration volume in a few domains, because the range of the real
exponential factors in the integrand varies considerably.

We studied the case of a smooth barrier with the form of a gaussian
ellipsoid in coordinate and momentum space. For values of $\rho$ and
$\varepsilon$ not much smaller than 1 a good precision was obtained. 
($\rho$ is the ratio between the lengths $a$ and $b$ of the ellipsoids
axes and $\varepsilon$ is the ratio between the length of the major axis
$a$ and the distance $L_1$ of the first source from the ellipsoid.)

Denoting by $n$ the distance of the second source in units of the major
axis, we found that the correction to the interaction potential along the
line joining the two sources and the barrier has the following form
(compare eq.s (\ref{e14}), (\ref{e28})): 
	\begin{eqnarray}
	U &=& U^0 + \gamma U' = \frac{1}{|{\bf x}_1 - {\bf x}_2|}
	-\frac{(2\pi)^{11}}{\sqrt{\pi}} \, \frac{\gamma ab}{L_1} 
	\, F(\varepsilon,\rho,n) \nonumber \\
	&=& \frac{1}{|{\bf x}_1 - {\bf x}_2|} \left[
	1 -\frac{(2\pi)^{11}}{\sqrt{\pi}} \gamma ab (1+n\varepsilon)
	F(\varepsilon,\rho,n) \right]. \nonumber
\end{eqnarray}
	The function $F$ depends very weakly on $\rho$ and $\varepsilon$.
Its dependence on $n$ is displayed in Fig.s 1-3 and shows an exponential
decay followed by a constant tail.

The behavior summarized above is interesting in itself, being the result
of a sort of ``tunneling" of the scalar field through a region where it is
constrained or has imaginary mass. We have seen that the local imaginary
mass term affects the propagation of the field also outside the region
$\Omega$ where it has support. This feature is easily understood from the
physical point of view; we gave here a method for its quantitative
evaluation.

\medskip

{\bf Acknowledgment} - This work has been partially supported by the
A.S.P., Associazione per lo Sviluppo Scientifico e Tecnologico del
Piemonte, Turin, Italy. 

\vskip 2 cm

\noindent
{\Large \bf Appendix: Proof of the expressions for $G'$, $U'$.}

\bigskip

We give here the proof of eq.s (\ref{e13}) and (\ref{e14}) of the 
main text. Expanding the square in (\ref{e11}) we obtain for $W[J]$
\begin{displaymath}
W[J]=\int d[\phi] \exp \left\{ -\int d^4x \, \left[(\partial \phi)^2 - 
2\xi \phi_0^2 J_\Omega(x) \phi^2(x)+ \xi J_\Omega(x) \phi^4(x)
+ \xi J_\Omega(x) \phi_0^4 \right] \right\}.
\end{displaymath}
The last term in the square bracket is constant with respect to $\phi(x)$
and its exponential can be factorized out of the functional integral.
In a first instance -- for weak
fields -- we can disregard the $\phi^4(x)$ term. We are then led to 
consider a 
quadratic functional integral, and the ``modified propagator"
$G(x,y)=\langle \phi(x) \phi(y) \rangle_J$, which by definition
satisfies the equation
\begin{equation}
\left[\partial^2_x + \gamma J_\Omega(x)\right] G(x,y) = -(2\pi)^4 
\delta^4(x-y), \label{e6}
\end{equation}
where $\gamma=2\xi \phi_0^2 >0$. Let us focus on the case when
$\phi_0^2=0$ inside the regions $\Omega_i$ and let us take the limit
$\phi_0 \to 0$ and $\xi \to \infty$ in such a way that $\gamma$ is finite
and very small, so that the term $\gamma J_\Omega(x)$ in eq.\ (\ref{e6})
constitutes only a small perturbation, compared to the kinetic term.
Then we can set
\begin{displaymath}
G(x,y)=G^0(x,y)+\gamma G'(x,y),
\end{displaymath}
where $G^0(x,y)$ is the propagator of the free scalar field, and we find
immediately that $G'(x,y)$ satisfies the equation
\begin{equation}
\partial^2_x G'(x,y) = - J_\Omega(x) G^0(x,y).
\label{spero}
\end{equation}

Unlike $G^0(x,y)$, in general $G'(x,y)$ will not depend only on $(x-y)$,
because the source breaks the translation invariance of the system.
In order to go to momentum space it will therefore be necessary to
consider
the Fourier transform of $G'(x,y)$ with respect to both arguments.
We define $\tilde{G}'(p,k)$ and $\tilde{J}_\Omega(p)$ as follows:
\begin{displaymath}
G'(x,y)=\int d^4p \int d^4k \, e^{ipx} e^{iky} \tilde{G}'(p,k)
\end{displaymath}
and
\begin{displaymath}
J_\Omega(x)=\int d^4p \, e^{ipx} \tilde{J}_\Omega(p), \qquad
G^0(x,y)=\int d^4k \, \frac{e^{-ik(x-y)}}{k^2} .
\end{displaymath}
The right hand side of (\ref{spero}) can be rewritten as
\begin{displaymath}
J_\Omega(x) G^0(x,y)=\int d^4p \int d^4k \, e^{ipx} \tilde{J}_\Omega(p)
\frac{e^{-ik(x-y)}}{k^2} =\int d^4k \int d^4p \, e^{iky} e^{ipx} 
\frac{\tilde{J}_\Omega(p+k)}{k^2}
\end{displaymath}
and we obtain the following algebraic equation for the double Fourier
transform of the first order correction to the propagator:
\begin{displaymath}
p^2 \tilde{G}'(p,k)=\frac{\tilde{J}_\Omega(p+k)}{k^2}.
\end{displaymath}
Transforming back, in conclusion we find eq.\ (\ref{e13}) of the main text, 
namely:
\begin{equation}
G'(x,y)=\int d^4p \int d^4k \, e^{ipx} e^{iky} \,
\frac{\tilde{J}_\Omega(p+k)}{k^2 p^2}.
\label{a13}
\end{equation}
Therefore, if we know the Fourier transform of the characteristic
function $J_\Omega$ of the spacetime region where the constraint is
imposed, we can in principle compute the leading order correction to
the field propagator and thus to $W[J]$.

It is known \cite{sym} that the vacuum-to-vacuum amplitude 
$W[J]=\langle 0^+ 
| 0^- \rangle_J$ of a field system in the presence of an external source $J$ 
is related to the logarithm of the systems' ground state energy:
\begin{displaymath}
E_0[J]= - T^{-1} \ln W[J],
\end{displaymath}
where the functional integral is supposed to be suitably normalized and
the source vanishes outside the temporal interval $[-T/2,+T/2]$, with 
$T$ eventually approaching infinity. (We use units in which $\hbar=c=1$.)

An interesting application of (\ref{a13}) occurs in the case when the
field $\phi(x)$ also interacts with $N$ static pointlike sources placed at
${\bf x}_1,{\bf x}_2...{\bf x}_N$. Namely, let us
add a further, linear coupling term $S_Q$ to the action of the system:
\begin{displaymath}
S_Q=\int d^4x \, Q(x)\phi(x), \qquad {\rm with} \qquad
Q(x)=\sum_{j=1}^N q_j \delta^3({\bf x}-{\bf x}_j) .
\end{displaymath}

The ground state energy of the system corresponds, up to a constant, 
to the static potential energy of the interaction of the sources through the 
field $\phi$. As before, it is obtained from the functional average of 
the interaction term, computed keeping the constraint into account:
\begin{equation}
E_0[J,Q] = U({\bf x}_1,...,{\bf x}_N)= 
-T^{-1} \ln \langle \exp \left\{-S_Q\right\} \rangle_J .
\label{a15}
\end{equation}
Expanding (\ref{a15}) one finds that to leading order in the 
$q_j$s, $U({\bf x}_1,...,{\bf x}_N)$ is given by a sum of
propagators integrated on time:
\begin{displaymath}
U({\bf x}_1,...,{\bf x}_N)= -T^{-1} \sum_{j,l=1}^N q_j q_l
\int dt_j \int dt_l \langle \phi(t_j,{\bf x}_j) \phi(t_l,{\bf x}_l) 
\rangle_J ,
\end{displaymath}
where $t_j,t_l \, \epsilon \, [-T/2,+T/2]$. Since the regions $\Omega_i$ are
infinitely elongated in the temporal direction the function
$\tilde{J}_\Omega(p+k)$ gets factorized as
\begin{equation}
\tilde{J}_\Omega(p+k)=(2\pi)^4\delta(p_0+k_0) \tilde{j}_\Omega({\bf p}+{\bf 
k}). \label{a17}
\end{equation}

Clearly the potential is disturbed by the presence of the ``barriers"
$j_\Omega({\bf x})$. To first order in $\gamma$ we can write
\begin{displaymath}
U({\bf x}_1,...,{\bf x}_N)=U^0({\bf x}_1,...,{\bf x}_N)+
\gamma U'({\bf x}_1,...,{\bf x}_N)
\end{displaymath}
and taking into account eq.s (\ref{a13}), (\ref{a17}) we find 
\begin{eqnarray}
U'({\bf x}_1,...,{\bf x}_N) & = & - T^{-1} \sum_{j,l=1}^N q_j q_l
\int dt_j \int dt_l \, G'(x_j,x_l) = \nonumber \\
& = & -(2\pi)^4 T^{-1} \sum_{j,l=1}^N q_j q_l \int dt_j \int dt_l 
\int d^4p \int d^4k \, \frac{e^{ipx_j+ikx_l} \tilde{J}_\Omega(p+k)}{k^2 p^2}= 
\nonumber \\
& = & -(2\pi)^8 T^{-1} \sum_{j,l=1}^N q_j q_l \int dt_j \int dt_l 
\int d^4p \int d{\bf k} \, \frac{e^{ip_0(t_j-t_l)+i{\bf px}_j + i{\bf kx}_l}
\tilde{j}_\Omega({\bf p}+{\bf k})}{(p_0^2+ {\bf k}^2)(p_0^2+ {\bf p^2})}.
\nonumber
\end{eqnarray}
Changing to variables $t=t_j-t_l$ and $s=t_j+t_l$ and integrating
we finally obtain the 
contribution of the perturbation to the static potential energy (eq.\ 
(\ref{e14}) of the main text):
	\begin{displaymath}
U'({\bf x}_1,...,{\bf x}_N)= -(2\pi)^8
\sum_{j,l=1}^N q_j q_l \int d{\bf p} \int d{\bf k} \,
e^{i{\bf px}_j + i{\bf kx}_l} \, \frac{\tilde{j}_\Omega({\bf p}+{\bf k})}
{{\bf k}^2 {\bf p}^2}.
\end{displaymath}

\newpage

\begin{center}
\begin{table}
\begin{center}
\begin{tabular}{|cccccccccc|} \hline \hline
${\bf \varepsilon}$ & ${\bf  \rho}$ & ${\bf   K_1}$ & ${\bf   P_1}$ & 
${\bf   K_2}$ 
& ${\bf   P_2}$  & ${\bf  K_3}$ & ${\bf   P_3}$& ${\bf   K_4}$  &  ${\bf 
P_4}$ \\ \hline
0.1 &  0.3  & 12  &  10 &   80  &  10 &   12 &   10 &   80   & 10 \\
0.1 &  0.1  & 10  &  10 &   100 &  25 &   20 &   20 &   120  & 40 \\
0.1 &  0.032 & 10  &  10 &   600 &  20 &   20 &   60 &   600  & 60 \\
0.1 &  0.01 & 20  &  20 &   1500 &  30 &   20 &   200 &   1500 & 200 \\
0.032 & 0.032 & 10  &  10 &   2000 &  15 &   20 &   70 &   1800 & 80 \\
0.032 & 0.01 & 12 & 12 & 6000 & 15 & 15 & 180 & 6000 & 250 \\
0.01 &  0.032 & 10  &  10 &   7500 &  15 &   10 &   70 &   7500 & 90 \\
0.032 & 0.0032 & 15 &  15 &   18000 & 15 &   15 &   650 &   18000 & 700 \\
\hline \hline
\end{tabular}
\end{center}
\caption{Integration ranges in the four domains, for some values of 
$\varepsilon$, $\rho$.} 
\end{table}
\end{center}
.
\newpage

\begin{center}
\begin{table}
\begin{center}
\begin{tabular}{|cccc|} \hline \hline
& {\bf b} & {\bf m} & {\bf q}
\\ \hline 
$\varepsilon=0.1$ \ (Fig.\ 1) & 0.14 & 0.29 & 0.3
\\ 
with $n>2$ & 0.12 & 0.24 & 0.1
\\ \hline
$\varepsilon=0.05$ \ (Fig.\ 3) & 0.17 & 0.32 & 0.4
\\
with $n>2$ & 0.16 & 0.26 & 0.1
\\ \hline \hline

\end{tabular}
\end{center}
\caption{Results of the best fit $F(n)=\exp(-mn+q)+b$.}
\end{table}
\end{center}

.
\newpage

\noindent
{\bf FIGURE CAPTIONS.}

\medskip
\medskip
\noindent Fig.\ 1 - Dependence of $F(\varepsilon,\rho,n)$ on $n$, in the
range $n$ =1...15, for $\varepsilon=0.1$ and $\rho=0.3$. Errors are $\sim$
0.01.

\medskip
\medskip
\noindent Fig.\ 2 - Comparison of the values of $F(\varepsilon,\rho,n)$
for $\varepsilon=0.1$ and $\rho=0.3$ (white circles), $\rho=0.1$ (black
triangles) and $\rho=0.032$ (black circles). 

\medskip
\medskip
\noindent Fig.\ 3 - Same as in Fig.\ 1, for $\varepsilon=0.05$. 
Errors are larger, as shown.

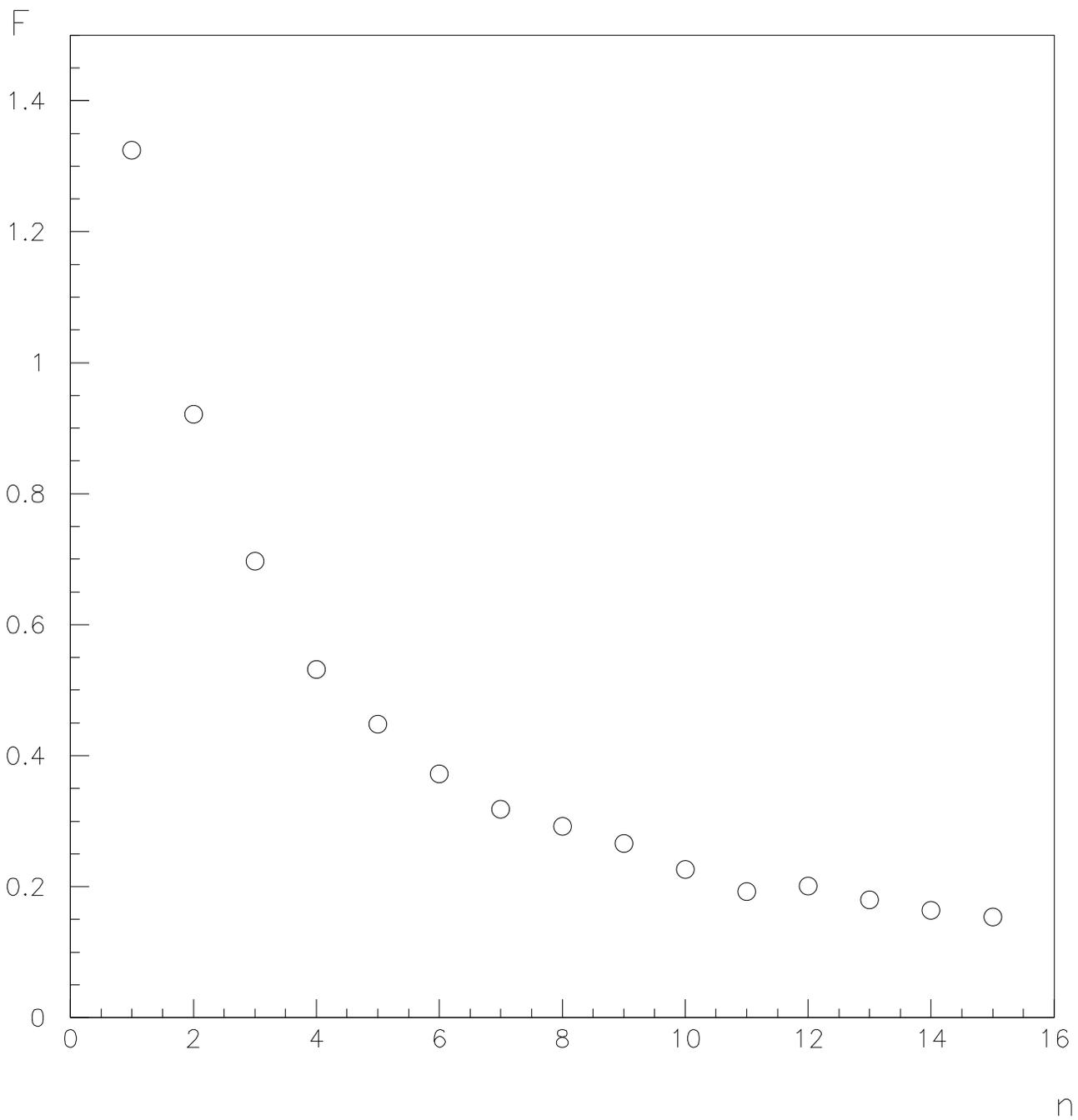
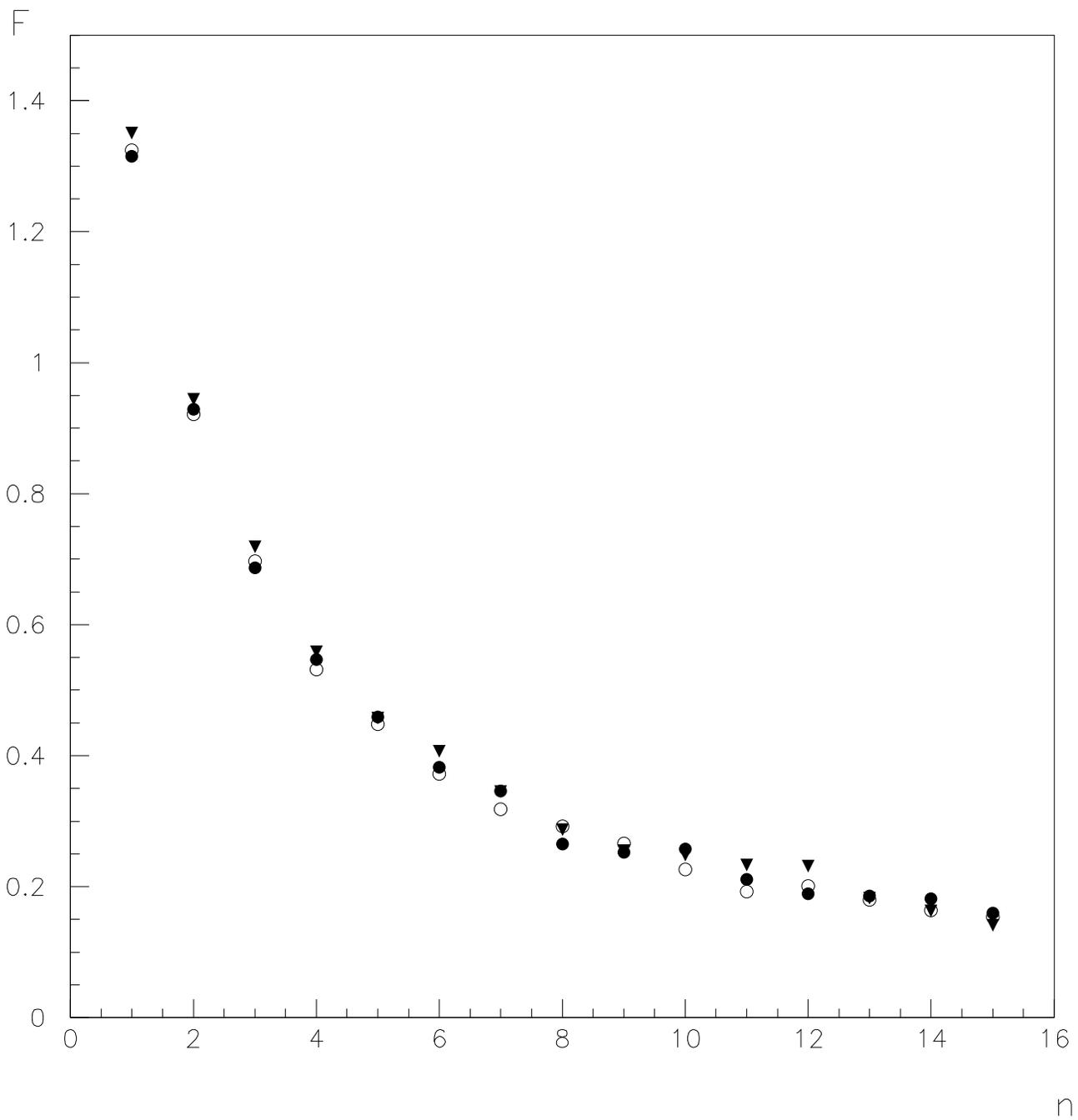
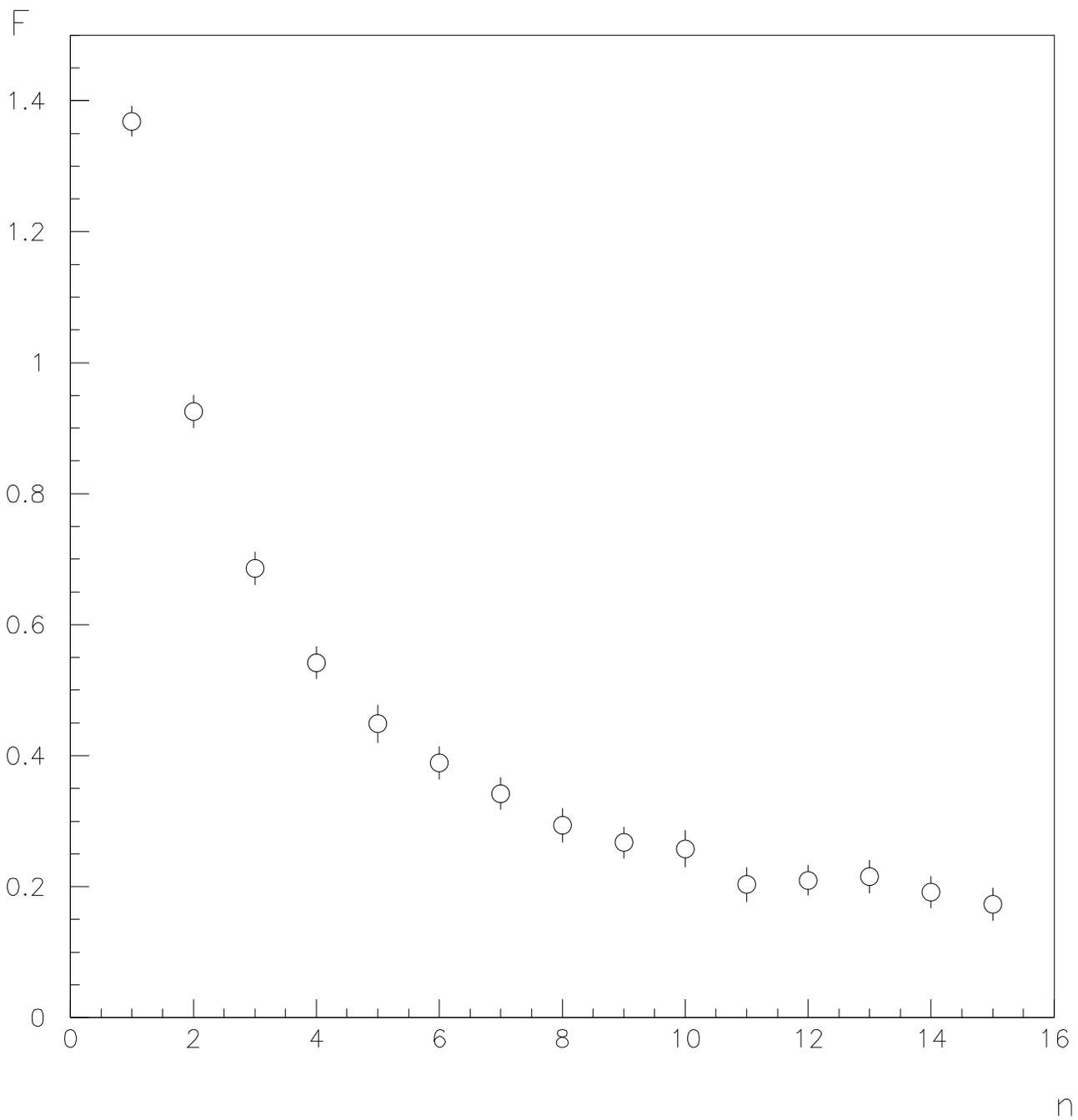
\end{document}